\documentclass{ws-procs9x6}
\usepackage{epsfig}

\begin{document}
\title{Charged Particle Multiplicities in Ultra-relativistic Au+Au and Cu+Cu Collisions}
\author{
B.B.Back\footnote{\uppercase{E}-mail: Back@anl.gov} \lowercase{ for the} PHOBOS C\lowercase{ollaboration}\\
B.Alver$^4$, B.B.Back$^1$, M.D.Baker$^2$, M.Ballintijn$^4$, D.S.Barton$^2$, R.R.Betts$^6$, A.A.Bickley$^7$,R.Bindel$^7$, W.Busza$^4$, A.Carroll$^2$, Z.Chai$^2$, V.Chetluru$^6$, M.P.Decowski$^4$, E.Garcia$^6$, T.Gburek$^3$, N.George$^2$, K.Gulbrandsen$^4$, C.Halliwell$^6$, J.Hamblen$^8$, I.Harnarine$^6$, M.Hauer$^2$, C.Henderson$^4$, D.J.Hofman$^6$, R.S.Hollis$^6$, R.Holynski$^3$, B.Holzman$^2$, A.Iordanova$^6$, E.Johnson$^8$, J.L.Kane$^4$, N.Khan$^8$, P.Kulinich$^4$, C.M.Kuo$^5$, W.Li$^4$, W.T.Lin$^5$, C.Loizides$^4$, S.Manly$^8$, A.C.Mignerey$^7$, R.Nouicer$^2$, A.Olszewski$^3$, R.Pak$^2$, C.Reed$^4$, E.Richardson$^7$, C.Roland$^4$, G.Roland$^4$, J.Sagerer$^6$, H.Seals$^2$, I.Sedykh$^2$, C.E.Smith$^6$, M.A.Stankiewicz$^2$, P.Steinberg$^2$, G.S.F.Stephans$^4$, A.Sukhanov$^2$, A.Szostak$^2$, M.B.Tonjes$^7$, A.Trzupek$^3$, C.Vale$^4$, G.J.vanNieuwenhuizen$^4$, S.S.Vaurynovich$^4$, R.Verdier$^4$, G.I.Veres$^4$, P.Walters$^8$, E.Wenger$^4$, D.Willhelm$^7$, F.L.H.Wolfs$^8$, B.Wosiek$^3$, K.Wozniak$^3$, S.Wyngaardt$^2$, B.Wyslouch$^4$}
\address{$^1$Argonne National Laboratory, Argonne, IL 60439, USA\\
$^2$Brookhaven National Laboratory, Upton, NY 11973, USA\\
$^3$Institute of Nuclear Physics PAN, Krakow, Poland\\
$^4$Massachusetts Institute of Technology, Cambridge, MA 02139, USA\\
$^5$National Central University, Chung-Li, Taiwan\\
$^6$University of Illinois at Chicago, Chicago, IL 60607, USA\\
$^7$University of Maryland, College Park, MD 20742, USA\\
$^8$University of Rochester, Rochester, NY 14627, USA}
\maketitle
\abstracts{The PHOBOS collaboration has carried out a systematic study of charged particle multiplicities in Cu+Cu and Au+Au collisions at the Relativistic Heavy-Ion Collider (RHIC) at Brookhaven National Laboratory. A unique feature of the PHOBOS detector is its ability to measure charged particles over a very wide angular range from 0.5$^\circ$ to 179.5$^\circ$corresponding to $|\eta|<$5.4. The general features of the charged particle multiplicity distributions as a function of pseudo-rapidity, collision energy and centrality, as well as system size, are discussed.
}

\section{Introduction}

The multiplicity of charged particles is a central observable in relativistic heavy-ion collisions, which provides information about the properties of the hot and dense fireball formed in such collisions. More detailed information about the system is embedded in identified particle spectra, but in most cases such spectra are not available over the full (pseudo)-rapidity range. The information that can be obtained from the non-identified charged particle measurements therefore provide unique opportunity to study the bulk properties of the system\cite{consequences}.   

\begin{figure}[hbt]
\centerline{\epsfig{file=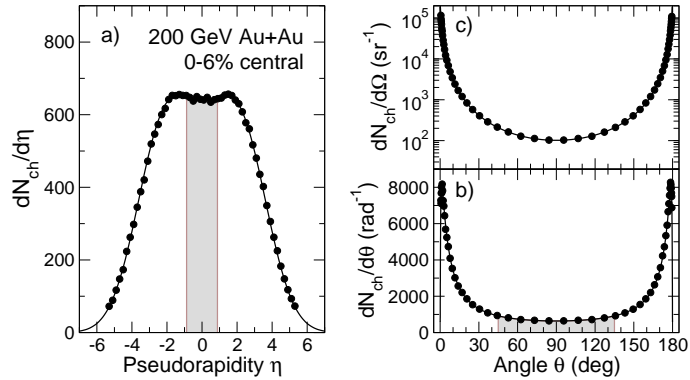,width=9cm}}
\caption{Illustration of the charged particle distribution for 0-6\% central 200 GeV $Au+Au$ collisions.
Panel a: The pseudorapidity density, $dN/d\eta$ is shown as a function of $\eta$. Panel b: The corresponding angular distribution $dN/d\theta$ is shown as a function of the angle $\theta$ relative to the beam axis. Panel c: same as for panel (b) but here $dN/d\Omega$ is shown. The shaded regions in panels (a) and (b) indicate the angular region where the transverse momentum $p_t$ exceeds the longitudinal momentum $p_{||}$.}
\label{shapes}
\end{figure}

Figure \ref{shapes}a shows the pseudo-rapidity distribution of charged particles emitted in $Au+Au$ collisions at $\sqrt{s_{\it NN}}$=200 GeV measured by the PHOBOS collaboration\cite{limfrag}. The typical mid-rapidity plateau extending over about four units of pseudorapidity, leads to a steep fall-off on either side. The solid curve represents a fit to the data using three Gaussians, one centered at $\eta$=0 and two located symmetrically on either side of mid-rapidity. This fit serves to estimate the total charged particle multiplicity by extrapolation into the unmeasured region. In this case we find that over 99\% of the charged particle yield falls into the PHOBOS acceptance region. The grey band encompasses the region, for which the transverse momentum exceeds the longitudinal momentum, $p_t>p_{||}$, which is most likely to reveal the signatures of the hot and dense fireball formed in the collision. Figs. \ref{shapes}b,c show, however, that the charged particle distribution is actually very forward-backward peaked in real space coordinates and that the mid-rapidity plateau corresponds to a minimum in the $dN/d\Omega$ distribution. 

\section{Multiplicity at mid-rapidity}

\begin{figure}[hbt]
\centerline{\epsfig{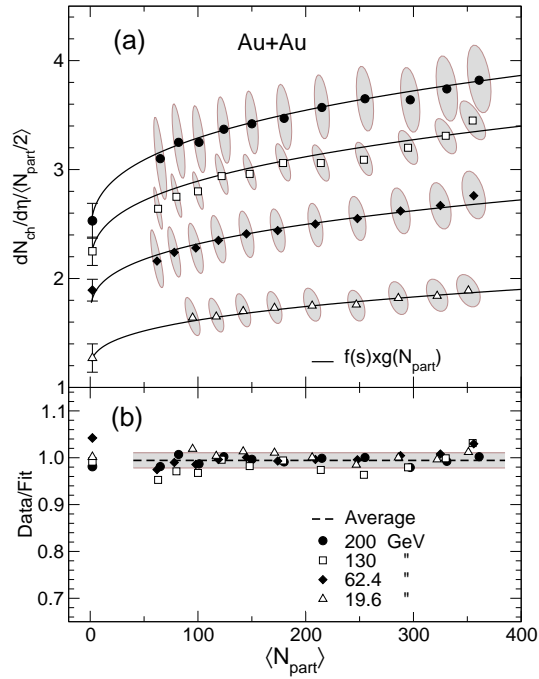}}
\caption{Illustration of charged particle distribution factorization for Au+Au collisions. Panel a: The pseudorapidity density, $dN/d\eta$ at $|\eta|<1$ is shown as a function of $N_{\it part}$ for $Au+Au$ and $pp/p\overline p$ collisions. Panel b: The ratio of the data to the fit function $f(s)\times g(N_{\it part})$ is seen to lie in a narrow band around unity.}
\label{factorization}
\end{figure}

The centrality dependence of the mid-rapidity multiplicity may be expressed in terms of the number of nucleons, $N_{\it part}$, of the initial $Au$ nuclei that participate in the collision. 
This variable is determined from the energy deposited in two scintillator paddle counters located at $3<|\eta|<4.5$ using Glauber model simulations\cite{Hollis}. In Fig.~\ref{factorization}a the measured charged-particle multiplicity at mid-rapidity\cite{200_20data,130_data,63_data}, normalized by $N_{\it part}/2$ in order to facilitate a comparison to elemental nucleon-nucleon collisions, is shown as a function of $N_{\it part}$ for collision energies of $\sqrt{s_{\it NN}}$=19.6, 62.4, 130, and 200 GeV. One observes that $dN_{\it ch}/d\eta/\langle N_{\it part}/2\rangle$ increases smoothly with both the centrality of the collision, expressed in terms of $\langle N_{\it part} \rangle$, and collision energy. For all centralities, the multiplicities  exceed those measured in $pp$\cite{Thome} and $p\overline p$\cite{Alner} collisions. In fact, the dependence on centrality and energy factorize to high accuracy as illustrated by the solid curves, which are given as the product two functions, {\it i.e.} $dN_{\it ch}/d\eta/\langle N_{\it part}/2\rangle = f(s) \times g(N_{\it part})$. The degree to which the factorization is valid is illustrated in Fig.~\ref{factorization}b, where the ratio of the data to the fit function is seen to deviate from unity by less than about 1\% (grey band). This factorization is surprising considering the significant changes in particle production mechanisms that are expected over the energy range of these data.

\section{System size dependence}

\begin{figure}[bt]
\centerline{\epsfig{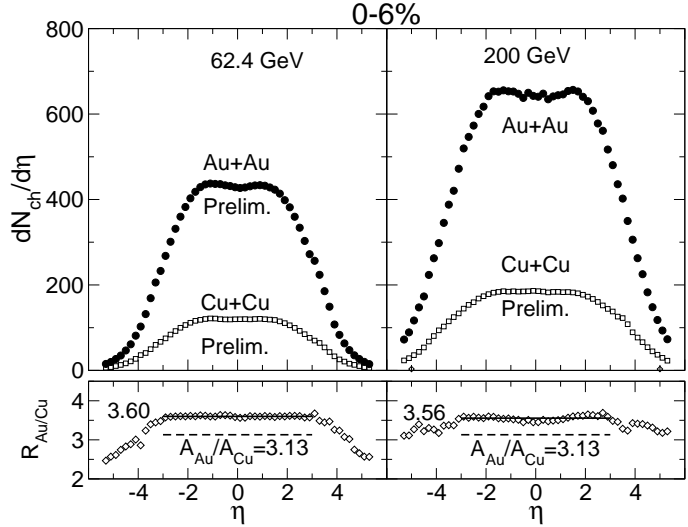}}
\caption{Comparison of charged particle multiplicity for $Au+Au$ and $Cu+Cu$ collisions at   62.4 GeV (left panels) and 200 GeV (right panels). The ratios of charged particle production density between $Au+Au$ and $Cu+Cu$ collisions are given as a function of pseudo-rapidity in the lower panels.}
\label{AuAu_CuCu}
\end{figure}

An alternative method of varying the size of the fireball is to collide ions of different size. Such a comparison has been carried out for central $Au+Au$ and $Cu+Cu$ collisions as illustrated in Fig.~\ref{AuAu_CuCu} at $\sqrt{s_{\it NN}}$=62.4 and 200 GeV. It is evident that similar shapes of the pseudorapidity distributions are found for both collision systems. Indeed, the lower panels of Fig.~\ref{AuAu_CuCu} show that the ratio of $dN_{\it ch}/d\eta$ between $Au+Au$ and $Cu+Cu$ collisions is essentially constant over the range $|\eta|<3$ at a value of 3.60 for 62.4 GeV and 3.56 for 200 GeV. These values are slightly larger than the ratio of the total number of nucleons available in the initial state, namely $A_{Au}/A_{Cu}$=3.13. 
Outside of $|\eta|<3$ the ratio falls off (steeply at 62.4 GeV and less so at 200 GeV) indicating that the high-$|\eta|$ tails fall off more steeply for $Au+Au$ than for $Cu+Cu$.

\section{Summary}

The PHOBOS experiment has a unique capability to measure nearly the full charged particle distribution as a function of pseudorapidity. At mid-rapidity the charged particle multiplicities have been measured for $Au+Au$ collisions as a function of both energy and centrality of the collisions. A surprising result is that the dependence on these two variables can be factorized to high accuracy ($\sim$ 1\%). A comparison of the pseudorapidity distributions for central $Au+Au$ and $Cu+Cu$ collisions shows that these exhibit the same shape over six units of pseudorapidity, but they differ somewhat in the tails of the distributions at high values of $|\eta|$.

\section*{Acknowledgments}
This work was partially supported by U.S. DOE grants DE-AC02-98CH10886, DE-FG02-93ER40802, DE-FC02-94ER40818, DE-FG02-94ER40865, DE-FG02-99ER41099, and W-31-109-ENG-38, by U.S. NSF grants 9603486, 0072204,  and 0245011, by Polish KBN grant 1-P03B-062-27(2004-2007), by NSC of Taiwan Contract NSC 89-2112-M-008-024, and by Hungarian OTKA grant (F 049823).

\end{document}